\documentstyle[12pt,psfig,epsf]{article}
\newcommand{\beq}{\begin{equation}}
\newcommand{\eeq}{\end{equation}}
\begin {document}

\hfill NTUA 63/97

\begin {center}
{\bf {\Large Three-dimensional lattice U(1) gauge-Higgs model}}
\end{center}
\begin {center}
{\bf {\Large  at low $m_H$}}

\vspace{1cm}

P.Dimopoulos, K.Farakos and G.Koutsoumbas

Physics Department

National Technical University, Athens

Zografou Campus, 157 80 Athens, GREECE

\vspace{2cm}

ABSTRACT
 
\end{center}

We study the non-compact version of the U(1) gauge-Higgs model 
in three dimensions for $m_H = 30 GeV.$ We found that, 
 using this formulation, rather modest lattices approach
 quite well the infinite volume behaviour.The phase transition is 
first order, as expected for this Higgs mass. 
The latent heat (in units of $T_{cr}^4$) is compatible with the 
predictions of the two-loop effective potential; it is an order of
magnitude less than the corresponding SU(2) value. 
The transition temperature and $<\varphi^* \varphi>$ in units of 
the critical temperature are also compatible with the perturbative results.

\newpage

\section{Introduction}

The main reason for the study of the three-dimensional 
gauge-Higgs system is its relation to the full 
$SU(2) \times U(1)$ Standard model at finite temperature. 
The latter has been studied extensively in recent years in 
connection  with the scenario of baryon violation at the 
electroweak scale during the evolution of the Early Universe.

It is well known that perturbation theory is not reliable for
the study of such models, because of severe infrared
divergencies. One promising approach has been to reduce the
four-dimensional model at finite temperature to an effective model
in three dimensions. 
This can be done if the couplings are small
and the temperature is much larger than any other mass scale in 
the theory [1, 2, 3]. The parameters of the reduced theory are 
related to the ones of the original model through perturbation 
theory. The reduced theory has some advantages over the
original one from the computational point of view [5, 6, 7]. It is 
super-renormalizable and yields transparent relations between
the (dimensionful) continuous parameters and the lattice ones. Moreover,
the number of mass scales is drastically reduced: (a) the scale $T,$ 
present in four dimensions is evidently absent, (b) one may also
integrate out the temporal component $A_0$ of the gauge field, 
so its mass scale $g T$ also
disappears. Thus there are two mass scales less and this 
reduces substantially the computer time needed to get reliable results. 

The model that has already been studied along these lines [5-8] has been based 
on $SU(2)$ with one complex Higgs doublet. (For work on the same model in 
asymmetric four-dimensional lattices, see [15].) 
The issues studied have 
been the order and the characteristics (critical temperature, latent
heat, surface tension, correlation lengths) of the phase transition, as
well as the reliability of perturbation theory deep in the broken 
phase for several values of the Higgs mass. It is interesting to
see how the above findings are affected by two characteristics of 
the study: the compactness of the gauge group and its non-abelian nature.

The abelian Higgs model has already been studied both in three and
four dimensions [14].The compact U(1) model has been studied on the 
lattice in [10, 11].An interesting aspect of the role of the abelian
character of the model would be to compare its latent heat against the 
one of the corresponding SU(2) theory.
 
We have chosen to concentrate on the non-compact model. To be exact,
only the kinetic term for the gauge field is written in the non-compact
form; for the kinetic term of the scalar field we use the compact 
formulation. This formalism has several advantages: 
\begin{itemize}
\item It follows closer
the continuum theory, so it should be
easier to approach the continuum limit. 
\item Our results show that with relatively small volumes one gets 
quite close to the thermodynamic limit.
\item The spurious
$U(1)$ monopoles, present in the compact formulation will not be
present any more. 
\item The non-compact version
is closer to the exact lattice form of the Landau-Ginzburg theory of 
superconductivity. The variable $x$ of our model, defined in the
text below, corresponds to $\kappa^2$ of the above  theory. We recall
that the the phase transition is of first order for type I 
superconductors ($\kappa < \frac{1}{\sqrt{2}}$), corresponding to small
Higgs mass; the converse holds for type II superconductors.
\end{itemize}

We have  concentrated on the phase transition line. We 
have chosen to fix the Higgs mass to a fixed value (30 GeV), 
g to 1/3, $m_W$ to 80.6 GeV and study the characteristics of 
the phase transition.

\section{Reduction of the four--dimensional theory to three dimensions}

The Lagrangian for the U(1) gauge--Higgs model in four dimensions is well
known:
\beq L_{4D} = \frac{1}{4}F_{\mu \nu}F_{\mu \nu} +|D_\mu \varphi |^2+
m^2\varphi ^* \varphi+\lambda (\varphi ^*\varphi)^2 \eeq

The action for this model at finite temperature is:
\beq
S[A_{\mu}(\tau ,\vec x) , \varphi (\tau ,\vec x)]=\int_{0}^{\beta } d\tau
\int d^3x [\frac{1}{4}F_{\mu\nu}F_{\mu\nu} +|D_{\mu }\varphi |^2+
m^2\varphi ^*\varphi+\lambda (\varphi ^* \varphi)^2], \eeq
where $\beta = 1/T.$

If the action is expressed in terms of Fourier components, the mass
terms are of the type:
\beq [(2\pi nT)^2 + (\vec k)^2]|A_\mu(n,\vec k)|^2 \eeq

\beq
[(2\pi nT)^2 + (\vec k)^2]|\varphi (n,\vec k)|^2,
\eeq
where $n=-\infty , \dots, \infty.$

At high temperatures T and energy scales less than $2\pi T$ the non--static modes
$A_\mu(n\neq 0,\vec k)$, $\varphi(n\neq 0,\vec k)$ are thus suppressed by
the factor $(2\pi n T)^2$ relative to the static $A_\mu(n=0,\vec k)$ and
$\varphi(n=0,\vec k)$ modes. The method of dimensional reduction consists
in integrating out the non--static modes in the action and deriving an
effective action [2, 3].

An important remark is that the mass of the adjoint Higgs field is of
order $g T,$ which is large compared to $g^2 T,$ the typical scale of the theory.
Thus one can go on one step further and integrate it out using perturbation
theory [5, 6, 7].

The effective action may then be written in the form:
\beq
S_{3D~eff}[A_{i}(\vec x),\varphi_{3}(\vec x)]=\int
d^3x[\frac{1}{4}F_{ij}F_{ij} +|D_{i}\varphi_3|^2+m_3^2\varphi_3^*
\varphi_3 +\lambda_3 (\varphi^*_3 \varphi_3)^2] \label{seff}
\eeq

The index 3 in (\ref{seff}) denotes the 3D character of
the theory. The relations between the 4D and 3D parameters are (up to 2
loops ):

\beq g_{3}^{2}=g^{2}(\mu)T, \eeq

\beq
\lambda_{3}=T(\lambda(\mu )+\frac{2}{(4\pi )^{2}}g^{4})
-\frac{g_{3}^{4}}{8\pi m_{D}},
\eeq

$$m_{3}^{2}(\mu _{3})=\frac{1}{4}g_{3}^{2}T+\frac{1}{3}(\lambda _{3}+
\frac{g_{3}^{4}}{8\pi m_{D}})T$$
$$+\frac{g_{3}^{2}}{16\pi ^{2}}
(-\frac{8}{9}g_{3}^{2}+\frac{2}{3}(\lambda _{3}+\frac{g_{3}^{4}}{8\pi m_{D}}))
-\frac{1}{2}m_{H}^{2}$$
\beq
+\frac{f_{2m}}{16\pi ^2} \log
(\frac{3T}{\mu_3}+c)-\frac{g_{3}^{2}m_{D}}{4\pi }-\frac{g_{3}^{4}}{8\pi
^{2}}(\! \log \frac{\mu_{3}}{2m_{D}}+\frac{1}{2}),
\eeq

\beq
m_{D}^{2}=\frac{1}{3}g^{2}(\mu )T^{2}.
\eeq

We note that $f_{2m}=-4g_{3}^{4} +8\lambda _{3} g_{3}^{2} -8\lambda _{3}^{2}$ 
and $c=-0.348725$ [5, 6, 7].
 
The couplings $g_{3}^{2}$, $\lambda_{3}$ of the three-dimensional theory
are renormalization group invariant because the theory is
supernormalisable. The mass parameter $m_{3}^{2}$ contains a linear and a
logarithmic divergence.

It is convenient to use the new set of parameters $(g_3^2,x,y)$ rather than the set
$(g_{3}^{2},\lambda_{3},m_{3}^{2}).$ $x,~y$ are defined by [8]: \beq
x=\frac{\lambda_{3}}{g_{3}^{2}} \eeq
\beq y=\frac{m_{3}^{2}(g_{3}^{2})}{g_{3}^{4}} \eeq

It is evident that $x$ is just proportional to the ratio of the squares of
the scalar and vector masses; on the other hand, $y$ is related to the temperature.
The parameters $x,~y$ can be expressed in terms of the four-dimensional
parameters as follows [10, 11]:
\beq
x=\frac{1}{2}\frac{m_{H}^{2}}{m_{W}^{2}}-\frac{\sqrt{3}\ g}{8\pi }
\label{mass}
\eeq

$$y=\frac{1}{4g^{2}}+\frac{1}{3g^{2}}(x+\frac{\sqrt{3}\ g}{8\pi })$$
$$+\frac{1}{16\pi ^{2}}
(-\frac{8}{9}+\frac{2}{3}(x+\frac{\sqrt{3}\ g}{8\pi }))-\frac{1}{4\pi \sqrt{3}\ g}$$
$$-\frac{1}{8\pi^2}(\! \log \frac{3 \sqrt{3}}{2g}+c+\frac{1}{2})-
\frac{m_{H}^{2}}{2 g^4 T^2}$$
\beq
+\frac{1}{16 \pi^2}(-4+8x-8x^2)(\! \log \frac{3}{g^2}+c) \label{T}
\eeq

\section{The lattice action}

Discretizing the continuum action (\ref{seff}) we get:

$$S=\beta _{g}\sum _{x}\sum _{0<i<j} F_{ij}^{2}+\beta _{h}\sum _{x}\sum
_{0<i}[ \varphi ^*(x)\varphi (x)-\varphi ^*(x)U_{i}(x)\varphi (x+\hat i)]$$
\beq
+\sum _{x}[(1-2\beta _{R}-3\beta _{h})\varphi ^*(x)\varphi (x)+\beta
_{R}(\varphi ^*(x)\varphi (x))^{2}], \label{lattact} \eeq

where $F_{ij}=\Delta _{i}^f A_{j}(x)-\Delta _{j}^f A_{i}(x),
~~U_{i}(x)=e^{iA_{i}(x)}.$

Notice that we use the non--compact version for the gauge field as explained 
in the introduction. The na\"{i}ve continuum limit corresponds to the values: 
$\beta_{g}= \infty,$ $\beta_{h}=\frac{1}{3},$ $\beta_{R}= 0.$

The lattice parameters and the (three-dimensional) continuum ones are related 
as follows \cite{laine}:
\beq
\beta_{g}=\frac{1}{ag_{3}^{2}}
\eeq

\beq
\beta_{R}=\frac{x\beta _{h}^{2}}{4\beta _{g}} \label{betar}
\eeq

$$2\beta _{g}^{2}(\frac{1}{\beta _{h}}-3-\frac{2\beta _R}{\beta
_{h}})=y-(2+4x)\frac{\Sigma \beta _{g}}{4\pi }$$
\beq
-\frac{1}{16\pi ^{2}}[(-4+8x-8x^{2})(\! \log 6\beta _{g}+0.09)-1.1+4.6x].
\label{y}
\eeq

We note that $\Sigma=3.176$ at the scale $\mu _{3}=g_{3}^{2}.$
 
\section{The algorithm}

We used the Metropolis algorithm for the updating of both the gauge and
the Higgs field. It is known that the scalar fields have much longer
autocorrelation times than the gauge fields. Thus, special care must be
taken to increase the efficiency of the updating for the Higgs field. We
made the following additions to the Metropolis updating procedure [8]:

{\bf a) Global radial update:} We update the radial part of the Higgs field
by multiplying it by the same factor at all sites: $R(\vec x)
\rightarrow e^{\xi }R(\vec x),$ where $\xi \in [-\varepsilon ,
\varepsilon]$ is randomly chosen. The quantity $\varepsilon$ is adjusted
such that the acceptance rate is kept between 0.6 and 0.7. The probability
for the updating is $P(\xi )=$ min$\{1,\exp (2V\xi -\Delta S(\xi)) \}$ where
$\Delta S(\xi )$ is the change in action, while the $2V\xi$ term comes
from the change in the measure.

{\bf b) Higgs field overrelaxation:} We write the Higgs potential at 
$\vec x$ in the form: \beq V(\varphi (\vec x))=-{\bf a} \cdot {\bf F} +
R^{2}(\vec x)+\beta _{h}(R^{2}(\vec x)-1)^{2} \eeq

where

$${\bf a} \equiv \left( \begin{array}{c} R(\vec x) \cos \chi (\vec x)\\
                                   R(\vec x) \sin \chi (\vec x)
                   \end{array}
                               \right),$$

$${\bf F} \equiv \left( \begin{array}{c} \beta_h \sum_i R(\vec x+ \hat i)\cos
(\chi (\vec x+\hat i)+\theta(\vec x)) \\
 \beta_h \sum_i R(\vec x+\hat i)\sin (\chi (\vec x+\hat i)+\theta(\vec x))
\end{array}
\right).$$

We can perform the change of variables:$({\bf a},{\bf F}) \rightarrow  
(X,F,{\bf Y})$ ,where
\beq
F \equiv |{\bf F}|,~~~ {\bf f} \equiv \frac{{\bf F}}{\sqrt{F_1^2 + F_2^2}},~~~
X \equiv {\bf a} \cdot {\bf f},~~~{\bf Y} \equiv {\bf a} - X {\bf f}.
\eeq

The potential may be rewritten in terms of the new variables:
\beq
\bar V(X,F, {\bf Y})=-XF +(1+2\beta _{R}({\bf Y}^{2}-1)) X^{2}
+{\bf Y}^{2}(1-2\beta_{R})+\beta _{R}(X^{4}+{\bf Y}^{4}).
\eeq

The updating of ${\bf Y}$ is done simply by the reflection:
\beq
{\bf Y} \rightarrow {\bf Y}'= -{\bf Y}.
\eeq

The updating of X is performed by solving the equation:
\beq
 (\frac{\partial \bar V(X',F',{\bf Y'})}{\partial X'})^{-1}
 \exp(-\bar V(X',F',{\bf Y'}))
=(\frac{\partial \bar V(X,F,{\bf Y})}{\partial X})^{-1} \exp(-\bar V(X,F,{\bf Y})).
\eeq

The change $X \rightarrow X'$ is accepted with probability:
 $P(X')=$ min$\{P_0,1\},$ where $P_0 \equiv
 \frac{\partial \bar V(X,F,{\bf Y})}{\partial X}/\frac{\partial \bar
 V(X',F',{\bf Y'})}{\partial X'}$.

\section{Results}

For our Monte--Carlo simulations we used cubic lattices with volumes
$V=$ $12^3,$ $~16^3$,$~24^3$. For each volume we performed 60000 to 110000
thermalization sweeps and 70000 to 120000 measurements. We have 
set the value of $x$ equal to 0.0463. According to the
relation (\ref{mass}), using $m_W=80.6 GeV$ and $g=\frac{1}{3},$ this value of
$x$ corresponds to a Higgs field mass $m_{H}=30 GeV.$ We used two values
for $\beta _g$, namely $\beta_g=4$ and $\beta_g=8$. For each value of
$\beta_h$ we use the relation (\ref{betar}) to determine the corresponding
$\beta_R.$ This value of $x$ has been used in references [10, 11] in the
study of the compact U(1) model, so we use the same value to facilitate
comparison. The two models should be close for large values of
$\beta_g,$ where the compact formulation probably approaches the
non-compact one. The phase transition is expected to be of first order,
since the mass of the scalar field is safely low.

We used four quantities to locate the phase transition points:

\begin{enumerate}

\item The distribution $N(E_{link})$ of $E_{link}.$

\item The susceptibility of $E_{link} \equiv \frac{1}{3V}
\sum_{x,i} \Omega^*(x)U_{i}(x) \Omega(x+i)$ (we have set $\varphi(x) \equiv
R(x) e^{i \chi(x)} \equiv R(x) \Omega(x)$ ):
$$S(E_{link}) \equiv V (<(E_{link})^{2}>-<E_{link}>^{2}).$$

\item The susceptibility of $R2 \equiv \frac{1}{V} \sum_x R^2(x):$
$$S(R2) \equiv V (<(R2)^2>-<R2>^2).$$

\item The Binder cumulant of $E_{link}$:
$$ C(E_{link})=1-\frac{<(E_{link})^4>}{3 <(E_{link})^2>^2}.$$

\end{enumerate}

We have searched for the (pseudocritical) $\beta_h$ values
 yielding (a) equal heights of the
two peaks of the distribution $N(E_{link}),$ (b) the maxima of the
quantities $S(E_{link}),~S(R2)$ and (c) the minima of the cumulant
$C(E_{link}).$ Of course,
the values $\beta_h^*(A,V)$ found using each of the above four quantities, 
depend on the specific quantity (denoted by A) which has been employed,
as well as on the volume V.  
While searching, we have made use of the Ferrenberg-Swendsen reweighting
technique [4] to find the pseudocritical $\beta_h$ for the volume $24^3$.

In figure 1 we show an example of the distribution of $E_{link}$ in a
$16^3$ lattice for $\beta_g = 8$ and three values of $\beta_h:$ the
pseudocritical one (0.337000), one somewhat smaller and one somewhat 
larger than this value.
This is just to illustrate the way in which the equal height criterion for
the critical point has worked. It is clear from the figure that the
pseudocritical $\beta_h$ yields two maxima of equal height in the distribution.
For the ``small" $\beta_h$ the peak in the region of small values of
$E_{link}$ is more pronounced, while for the ``large" $\beta_h$ it is the
other way around. The picture of the two well separated peaks at
criticality is the signature of a first order phase transition; the two
peaks correspond to the two coexisting metastable states.

%%%%%%%%%%%%%%%%%%%%%%%%%%%%%%%%%%%%%%%%%%%%%%%%%%%%%%%%%%%%%%%%
\begin{figure}
\centerline{\hbox{\psfig{figure=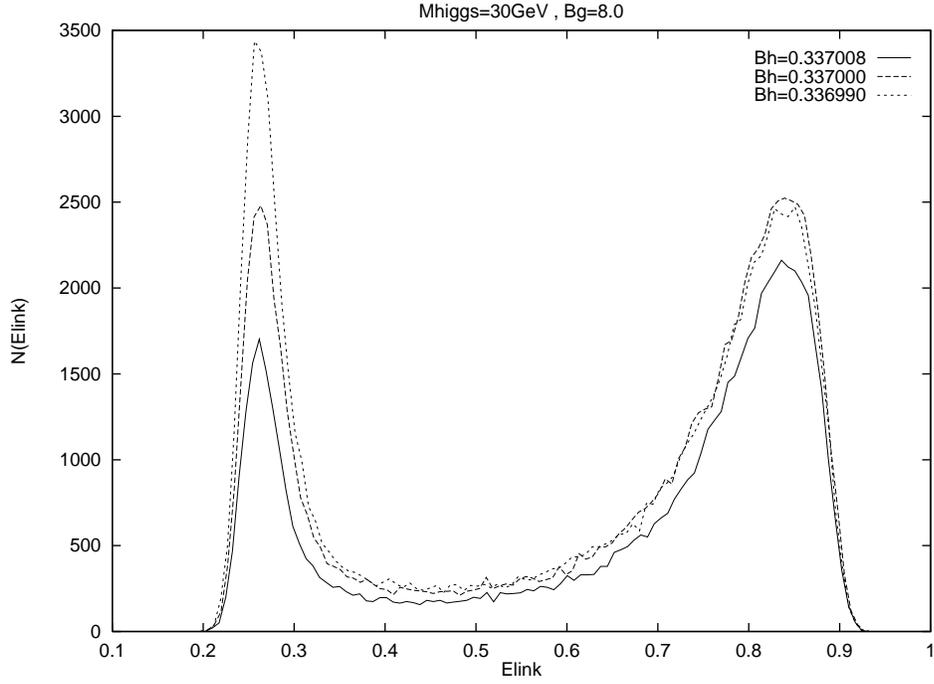,height=9cm,angle=-90}}}
\caption[f1]{Distribution of $E_{link}$}
\label{f1}
\end{figure}
%%%%%%%%%%%%%%%%%%%%%%%%%%%%%%%%%%%%%%%%%%%%%%%%%%%%%%%%%%%%%%%%

In figures 2 and 3 we depict the behaviour of the susceptibilities
$S(E_{link})$ and $S(R2)$ for $\beta_g = 4$ versus $\beta_h$ for three
lattice volumes. We have fitted curves through the data and show them in
the figures; for the $16^3$ and $24^3$ lattice volumes we also give 
the actual measurements.
It is evident that the curves represent the data quite
nicely. To calculate the error bars we first found the integrated
autocorrelation times $\tau_{int}(A)$ for the relevant quantities $A$ and
constructed samples of data separated by a number of steps greater than
$\tau_{int}(A).$ The errors have been calculated by the Jackknife method,
using the samples constructed according to the procedure just described. We
observe that the peak values for the susceptibilities increase almost
linearly with the volume in both cases, which is evidence for a first order
phase transition.
%%%%%%%%%%%%%%%%%%%%%%%%%%%%%%%%%%%%%%%%%%%%%%%%%%%%%%%%%%%%%%%%
\begin{figure}
\centerline{\hbox{\psfig{figure=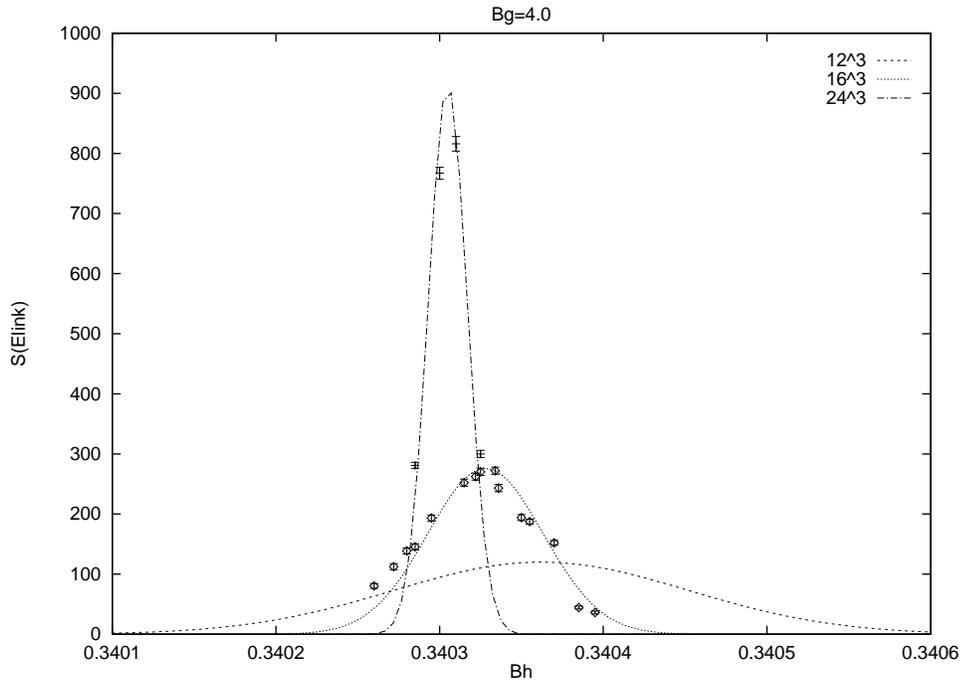,height=9cm,angle=-90}}}
\caption[fig2a]{Susceptibility of $E_{link}$}
\label{fig2a}
\end{figure}
%%%%%%%%%%%%%%%%%%%%%%%%%%%%%%%%%%%%%%%%%%%%%%%%%%%%%%%%%%%%%%%%
%%%%%%%%%%%%%%%%%%%%%%%%%%%%%%%%%%%%%%%%%%%%%%%%%%%%%%%%%%%%%%%%
\begin{figure}
\centerline{\hbox{\psfig{figure=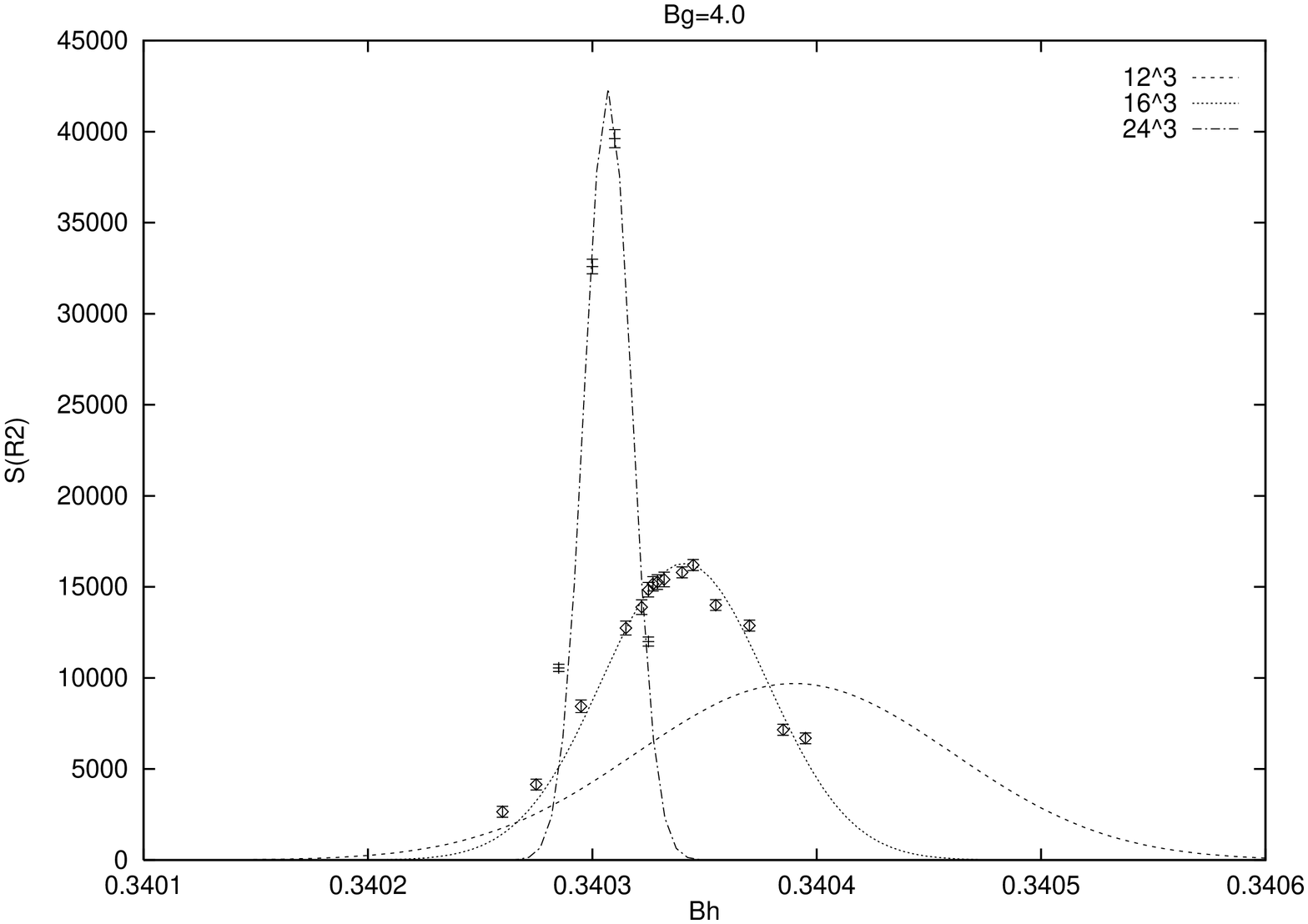,height=9cm,angle=-90}}}
\caption[fig2b]{Susceptibility of $R2$}
\label{fig2b}
\end{figure}
%%%%%%%%%%%%%%%%%%%%%%%%%%%%%%%%%%%%%%%%%%%%%%%%%%%%%%%%%%%%%%%%
In figure 4 we depict the behaviour of the Binder cumulant $C(E_{link})$ at
$\beta_g = 8$ for three lattice sizes. We again show the real measurements
for $16^3$ and $24^3$  only and just give the fitted curves for $12^3.$
 The error bars have been calculated by the Jackknife
method [13], in the same way as in the case of the susceptibilities. The volume
dependence of the cumulants is again characteristic of a first order phase
transition.
%%%%%%%%%%%%%%%%%%%%%%%%%%%%%%%%%%%%%%%%%%%%%%%%%%%%%%%%%%%%%%%%
\begin{figure}
\centerline{\hbox{\psfig{figure=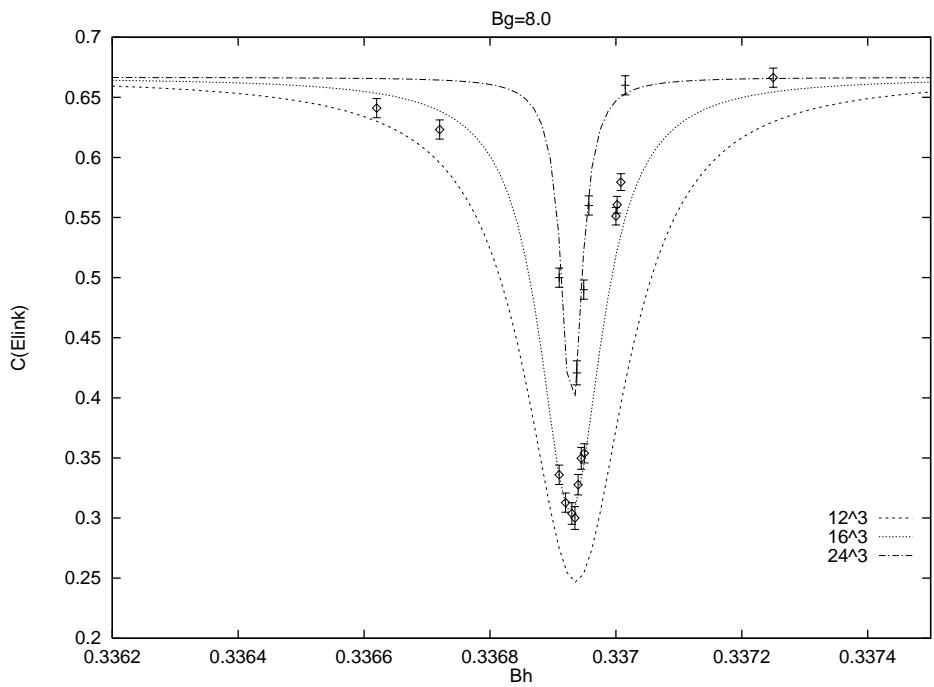,height=9cm,angle=-90}}}
\caption[fig3]{Binder cumulant of $E_{link}$}
\label{fig3}
\end{figure}
%%%%%%%%%%%%%%%%%%%%%%%%%%%%%%%%%%%%%%%%%%%%%%%%%%%%%%%%%%%%%%%%

The use of finite lattices is the reason why the $\beta_h^*(A,~V)$ values 
that we have found employing
the various criteria are slightly different. Thus, one should extrapolate
these values to infinite volume. We have adopted the ansatz: 
$$\beta_h^*(A,~V) = \beta_h^{cr}(\infty) +\frac{c(A)}{V},$$ 
The constant $c(A)$ is
expected to depend on the quantity $A$, while the extrapolated value
$\beta _{h}^{cr}(\infty )$ should not depend on $A$; that is, the infinite
volume extrapolation for the critical point should not depend on the
quantity used.

Figures 5 and 6 deal with the extrapolation to infinite volume for
$\beta_g=4$ and $\beta_g=8$ respectively. They contain the data for the
pseudocritical $\beta_h^*(A,~V)$ values obtained from the various
quantities $A$ versus the inverse lattice volume, along with the linear
fits to the data.The error bars in $\beta _h$ have been found from the
statistical error of the values of the quantities A at the critical 
point.  
We note in passing that, at
finite volumes, the smallest pseudocritical values are given by the
cumulant of $E_{link}$; then follow, in ascending order, the values given
from  the equal height, the susceptibility of
$E_{link}$ and the susceptibility of $R2.$This holds for both values of
$\beta _g.$
One may also observe that the infinite volume extrapolation is
almost independent from the specific quantity used: the differences at the
point $\frac{1}{V} = 0$ between the various extrapolated critical values
are less than $10^{-5}.$ To be specific, the critical values lie in the
interval (0.340295, 0.340298) for $\beta_g = 4$ and in (0.336932, 0.336940)
for $\beta_g = 8.$ In reference [10], treating the compact U(1)
model, the best pseudocritical value correesponding to our results 
has been obtained for $\beta_g=8$ for a $32^3$
lattice. From the relevant figure one may read out a pseudocritical value
about 0.3370, a result consistent with ours. These results suggest that the
non-compact formulation allows one to obtain similar results to the ones
of the compact formulation in a quite economical way.
%%%%%%%%%%%%%%%%%%%%%%%%%%%%%%%%%%%%%%%%%%%%%%%%%%%%%%%%%%%%%%%%
\begin{figure}
\centerline{\hbox{\psfig{figure=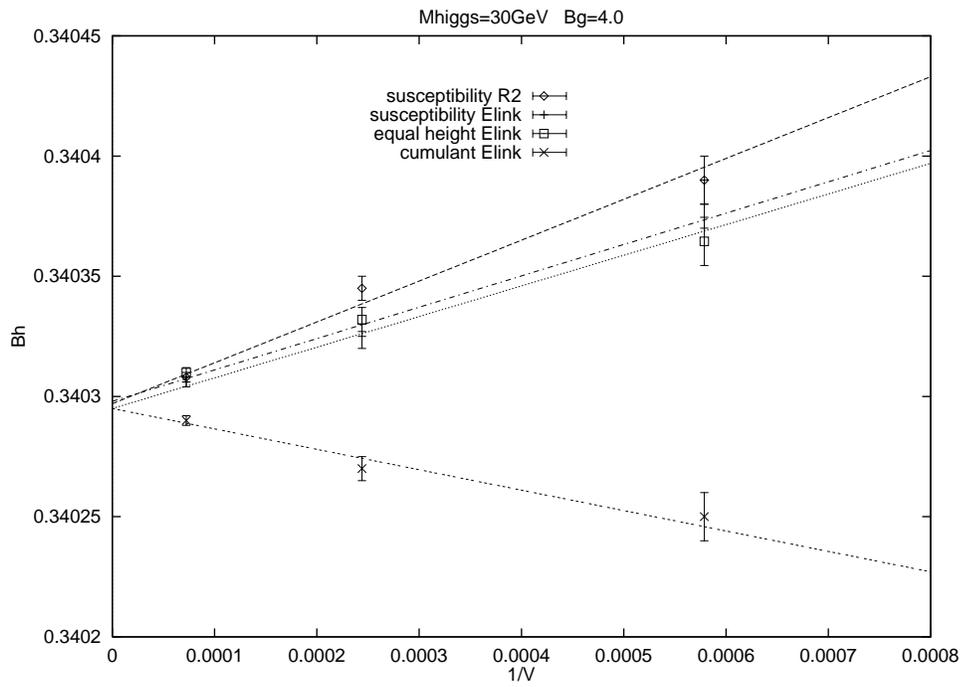,height=9cm,angle=-90}}}
\caption[fig4a]{Extrapolation for $\beta_g=4$}
\label{fig4a}
\end{figure}
%%%%%%%%%%%%%%%%%%%%%%%%%%%%%%%%%%%%%%%%%%%%%%%%%%%%%%%%%%%%%%%%

%%%%%%%%%%%%%%%%%%%%%%%%%%%%%%%%%%%%%%%%%%%%%%%%%%%%%%%%%%%%%%%%
\begin{figure}
\centerline{\hbox{\psfig{figure=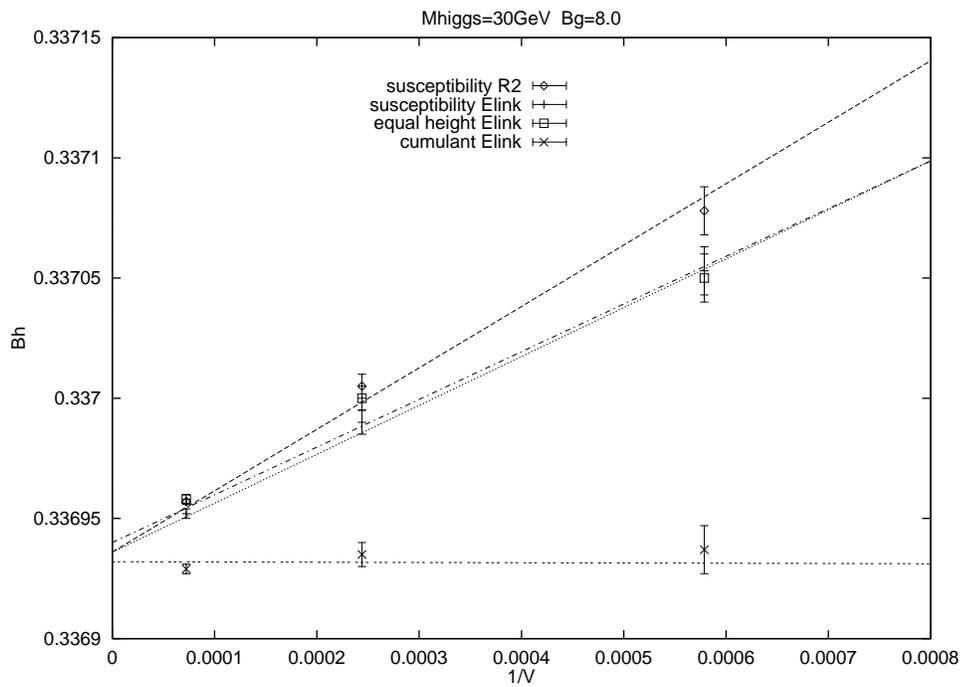,height=9cm,angle=-90}}}
\caption[fig4b]{Extrapolation for $\beta_g=8$}
\label{fig4b}
\end{figure}
%%%%%%%%%%%%%%%%%%%%%%%%%%%%%%%%%%%%%%%%%%%%%%%%%%%%%%%%%%%%%%%%
The next quantity we are going to deal with is the critical temperature.
It may be determined by noting that, for each $\beta_g$, the quantity
$\beta_h^{cr}(\infty)$ yields $y_{cr}$ through equations (\ref{betar},
\ref{y}); then equation (\ref{T}) gives $T_{cr}.$ The results are to be
found in table 1. We mention that in 
reference [8] the critical temperature for the SU(2) model at 
$\beta _g=8$ and $m_{H}=35GeV$ has been found $94.181GeV.$  
No lattice result is reported for this
quantity in the paper \cite{karjal2} on compact U(1), but there is the result
$148.83 GeV$ at $m_H=35 GeV$ from the perturbative effective potential.
We will say more about this later on, but we remark at this point that
this value is quite close to ours.

Having found the critical temperature we estimated the latent heat, that
is the energy released in the transition. We have used the formula [8, 9]:
\beq \frac{L}{T_{cr}^{4}}=\frac{1}{2} \frac{M_H^2}{T_{cr}^{3}} g_3^2
\beta_h^{cr} \beta_g \Delta <R2>. \eeq

We note that $g_3^2 = g^2 T_{cr}.$ The quantity $\Delta <R2>$ is the
difference of the $R2$ expectation values between the phases. We have
measured the values of $\Delta <R2>$ from the $R2$ distributions for each
lattice volume at the three different pseudocritical values of
$\beta_h^*(A,V)$ [12] the quantities $A$ being the susceptibilities of
$E_{link}$ and $R2$ and the equal height signal of $N(E_{link})$. Figures 7
and 8 show these sets of three measurements versus the inverse volume of the
lattice for the cases $\beta_g = 4$ and $\beta_g = 8.$ The error bars are
due to the uncertainty of each pseudocritical value, as well as to
statistical dispersion; they turn out to be rather big, especially for the
smallest volume. The final values of $\Delta <R2>$, which have been used
for the calculation of $\frac{L}{T_{cr}^4}$ have been found from a linear
fit of these data with $\frac{1}{V}.$ The convergence of the three
straight lines to a common value in the limit of infinite volume is
fairly good. 
%%%%%%%%%%%%%%%%%%%%%%%%%%%%%%%%%%%%%%%%%%%%%%%%%%%%%%%%%%%%%%%%
\begin{figure}
\centerline{\hbox{\psfig{figure=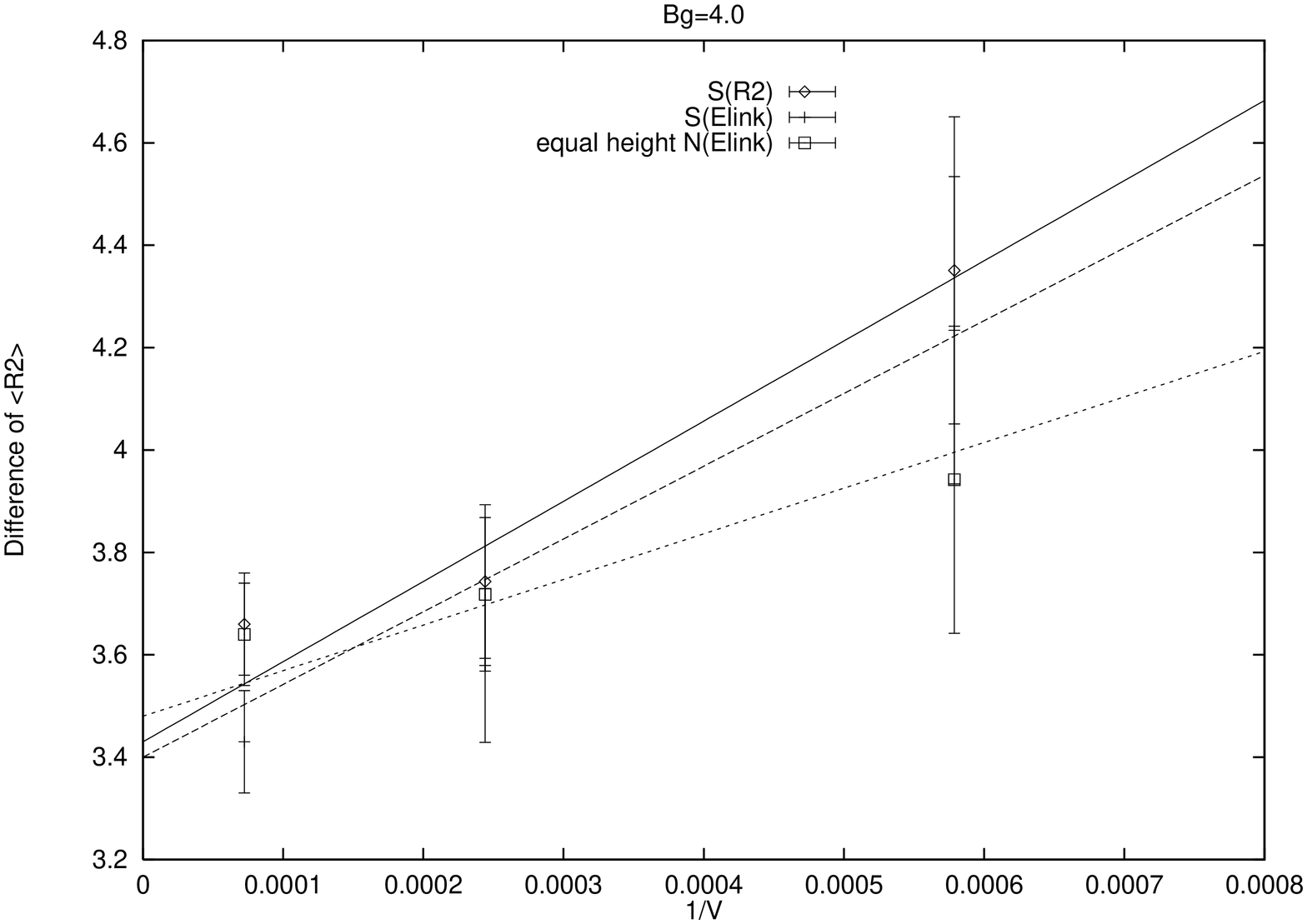,height=6cm,angle=-90}}}
\caption[fig5a]{Extrapolation for $\Delta <R2>$}
\label{fig5a}
\end{figure}
%%%%%%%%%%%%%%%%%%%%%%%%%%%%%%%%%%%%%%%%%%%%%%%%%%%%%%%%%%%%%%%%
%%%%%%%%%%%%%%%%%%%%%%%%%%%%%%%%%%%%%%%%%%%%%%%%%%%%%%%%%%%%%%%%
\begin{figure}
\centerline{\hbox{\psfig{figure=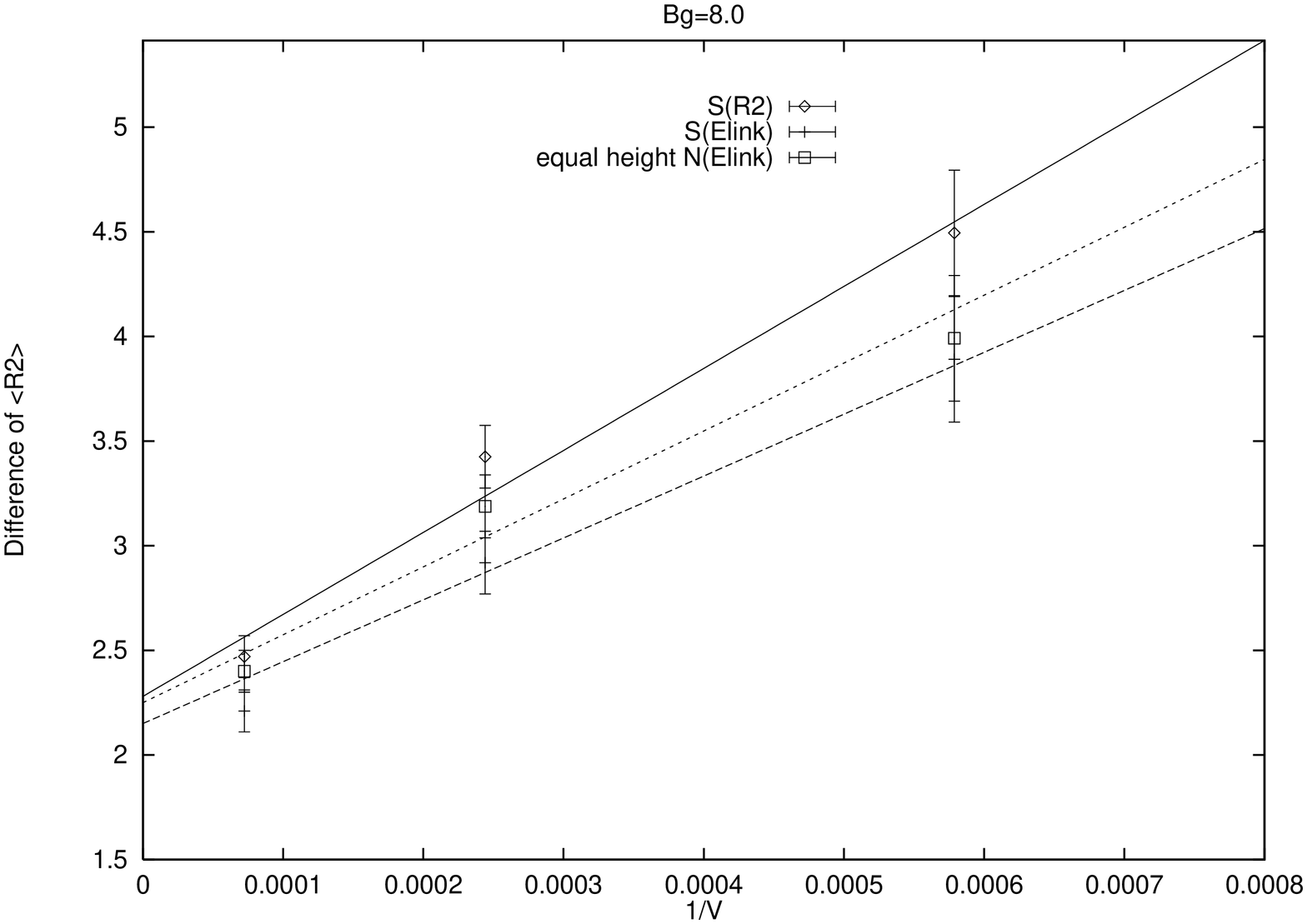,height=6cm,angle=-90}}}
\caption[fig5b]{Extrapolation for $\Delta <R2>$}
\label{fig5b}
\end{figure}
%%%%%%%%%%%%%%%%%%%%%%%%%%%%%%%%%%%%%%%%%%%%%%%%%%%%%%%%%%%%%%%%
The results for the quantity $\frac{L}{T_{cr}^4}$ can be found in table 1.
The corresponding results for the SU(2) case at $m_H=35GeV$ have been
reported to be $0.180 \pm 0.001$ in \cite{ger} and $0.256 \pm 0.008$
in \cite{klrs}. The former result has been obtained for $\beta_g=12,$
while for the latter an extrapolation to $\beta_g = \infty$ 
has been done. We
observe that our values of this quantity for U(1)
 are less than the one tenth of
the values for $SU(2).$ This permits one to
be sure that the $U(1)$ part of the Standard Model gauge
group plays only a secondary role in the scenario of the Electroweak
Phase Transition. The relative factor of ten is so big, that this
conclusion cannot be spoiled by the rather large errors. 

\[
%%%%%%%%%%%%%%%%%%%%%%
\begin{array}{||c|c|c|c||}
\hline
\multicolumn{4}{||c||}{Table~~1} \\ \hline
  \beta_g    &T_{cr}    &L/T_{cr}^4   &<\varphi^* \varphi>/T_{cr}   \\
             &          &             &          \\ \hline
 4           &131.50(3)    &0.0135(4)   & 0.255(6)          \\
 8           &131.18(14)   &0.0172(6)    & 0.308(10)           \\ \hline
 pert.       &132.64       &0.0150      & 0.285          \\ \hline
\end{array}
%%%%%%%%%%%%%%%%%%%%%%
\]

For the (3-dimensional) lattice quantity
 $<\varphi^* \varphi>/T_{cr},$ also appearing in table 1,
it is important that one subtracts the ``infinities" from the lattice
results. The relevant formula reads \cite{laine}:
$$
<\varphi^* \varphi> =\frac{1}{2} \beta_h \beta_g g_3^2 <R2>
-\frac{g_3^2 \beta_g \Sigma}{4 \pi} -\frac{g_3^2}{8 \pi^2}
\left[log 6 \beta_g +\zeta+\frac{\Sigma^2}{4}-\delta \right],
$$
where $\Sigma=3.176,~~\zeta=0.09$ and $\delta=1.94.$

\begin{figure}
\centerline{\hbox{\psfig{figure=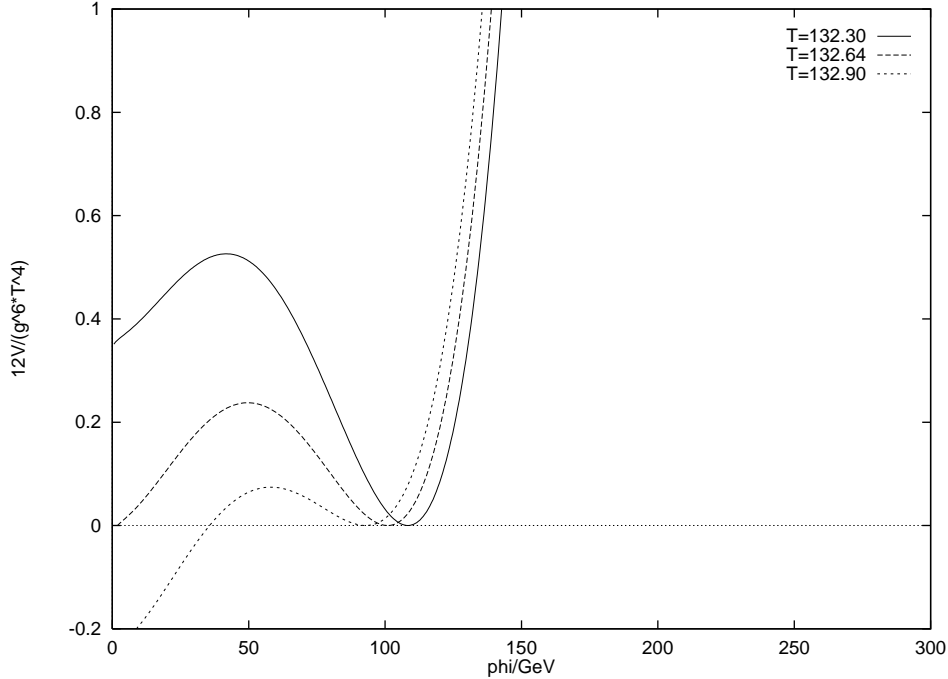,height=9cm,angle=-90}}}
\caption[eff]{2--loop effective potential.We use $phi=\varphi \sqrt{T}$}
\label{eff}
\end{figure}
 
One may use the $U(1)$ effective potential \cite{fkrs1} to determine the
critical temperature $T_{cr}$, as well as the quantity
$<\varphi^* \varphi>/T_{cr}$ and $\frac{L}{T_{cr}^4}$
 and compare with 
the corresponding quantities from the lattice. (We note that the
critical temperature is defined in perturbation theory by the equality of
the two minima of the potential.)
The perturbative predictions are also displayed in table 1.

Figure 9 depicts the two-loop effective potential versus the {\bf 
four-dimensional} scalar field (a) for the critical
temperature $T_{cr, pert},$ and (b) for two other neighbouring temperatures,
one corresponding to the symmetric phase and the other to the 
broken phase. 

In principle one should perform the extrapolation to large values of 
$\beta_g$. However one is not sure about the exact $\beta_g$ dependence
of the various quantities, so we postpone this until we get results 
for even bigger $\beta_g.$

We observe that the lattice $T_{cr}$ is smaller than the prediction
from perturbation theory and  decreases with
$\beta_g$ (in agreement with the SU(2) results for small $m_{H}$ [8]).
The other two quantities that we measured, 
 namely $\frac{L}{T_{cr}^4}$ and
$<\varphi^* \varphi>/T_{cr}$, increase with $\beta_g$ and their
values are compatible with the perturbative ones. 

{\bf Acknowledgements}

K.F. thanks M.Laine for a crucial observation concerning equation (17).
This work has been supported in part by the PENED 95 Program, No 1170 of
the Greek General Secretariat of Research and Technology. Thanks are also
due to the Computer Center of the NTUA and the Silicon Graphics Hellas Ltd
for the computer time allocated.

\end{document}